\documentclass{l4dc2025}

\usepackage{hyperref}
\usepackage{algorithm}
\usepackage{caption}

\title[Finding the Human Optimum in Human-Machine Interaction Games]{A Learning Algorithm That Attains the Human Optimum in a Repeated Human-Machine Interaction Game}
\usepackage{times}

\newcommand{\wh}[1]{\widehat{#1}}

\coltauthor{\Name{Jason T. Isa} \Email{jisa@uw.edu}\\
 \Name{Lillian J. Ratliff} \Email{ratliffl@uw.edu}\\
 \Name{Samuel A. Burden} \Email{sburden@uw.edu}\\
 \addr University of Washington, Seattle, WA, United States}

\begin{document}

\maketitle

\begin{abstract}
When humans interact with learning-based control systems, a common goal is to minimize a cost function known only to the human.
For instance, an exoskeleton may adapt its assistance in an effort to minimize the human's metabolic cost-of-transport.
Conventional approaches to synthesizing the learning algorithm solve an inverse problem to infer the human's cost.
However, these problems can be ill-posed, hard to solve, or sensitive to problem data.
Here we show a game-theoretic learning algorithm that works solely by observing human actions to find the cost minimum, avoiding the need to solve an inverse problem.
We evaluate the performance of our algorithm in an extensive set of human subjects experiments, demonstrating consistent convergence to the minimum of a prescribed human cost function in scalar and multidimensional instantiations of the game.
We conclude by outlining future directions for theoretical and empirical extensions of our results.
\end{abstract}

\begin{keywords}%
  human-machine interaction, learning-based control, human subjects experiments 
\end{keywords}

%%%%%%%%%%%%%%%%%%%%%%%%%%%%%%%%%%%%%%%%%%%%%%%%%%%%%%%%%%%%%%%%%%%%%%%%

%%% Include any author-defined commands here.
         
\newcommand{\TODO}[1]{}

\newcommand{\mc}{\mathcal}

\newcommand{\vep}{\varepsilon}

\newcommand{\R}{\mathbb{R}}

\newcommand{\BR}{{\tt BR}}

%%%%%%%%%%%%%%%%%%%%%%%%%%%%%%%%%%%%%%%%%%%%%%%%%%%%%%%%%%%%%%%%%%%%%%%%

\section{Introduction}

Machines -- including robots and learning algorithms -- are increasingly moving from the confines of labs and factories to interact with humans in daily life~\citep{Hoc2000From,cannan2011human,Gorecky2014Human,Kun2018Human}. 
Adaptive machines provide an exciting potential to assist humans in everyday work and activities as tele- or co-robots~\citep{nikolaidis2017game}, interfaces between computers and the brain or body~\citep{perdikis2020brain, DeSantis2021}, and devices like exoskeletons or prosthetics~\citep{Felt2015Body, Zhang2017Human, slade2022personalizing}. 
But designing learning algorithms that can safely interact with humans and constantly adapt to a dynamic environment remains an open problem in robotics, neuroengineering, and machine learning~\citep{nikolaidis2017game, Recht2019, perdikis2020brain}. 

In this work, we embrace the broadly-held hypothesis that human behavior is governed by minimization of a cost function~\citep{VonNeumann1944,simmon1955,Todorov2002-sb}.
When this hypothesis holds, an assistive machine's goal is to find this cost's minimum.
We propose a new learning algorithm that yields convergence to the human's optimum in a repeated game formulation of the human-machine interaction. 
% sam: I'm not sure this is accurate
Rather than having prior knowledge of the human's cost function or solving an inverse problem to estimate it as in prior work~\citep{ng2000algorithms,Merel2013-jd,li2019differential}, our algorithm achieves this outcome solely through observations of the human's actions over repeated interactions.
This feature may be valuable in the context of the emerging body/human-in-the-loop optimization paradigm for assistive devices~\citep{Felt2015Body, Zhang2017Human, slade2022personalizing}, where the machine continually interacts with the human but does not have direct access to the human's cost function, e.g.\ metabolic energy consumption~\citep{Abram2022General} or other preferences~\citep{Ingraham2022Role}. 

Game theory is an established field concerned with strategic interactions between two or more decision-makers~\citep{VonNeumann1944}. 
Prior work has modeled human-machine interaction in a game theory framework, including~\citep{li2016framework,nikolaidis2017game,crandall2018cooperating,li2019differential,cao2020portfolio,march2021strategic,chasnov2023human,isa2024effect}.
The game we consider here is repeated continuously in time over continuous action spaces, distinguishing our setting from some prior work.
Moreover, we consider a special imperfect information set wherein both agents seek to minimize a shared cost but one agent (the human) has perfect knowledge of the cost whereas the other (the machine) can only observe actions of both players.
We furthermore assume there is no communication or collusion between the agents other than revealing their actions.
The machine's task is to find the cost minimum solely by playing actions or policies and observing the human's response.  
Although it may seem to be an impossible task, we in fact show that a simple learning algorithm yields convergence to the cost's minimum. 

The learning algorithm we propose is the first of its kind, yielding convergence to the minimum of a cost that is unknown from the machine's perspective. 
This algorithm may be particularly valuable in providing optimal assistance for a prosthetic or exoskeleton~\citep{Felt2015Body, Zhang2017Human, slade2022personalizing}, where the machine does not have direct access to the human's cost function, e.g.\ metabolic energy consumption~\citep{Abram2022General} or other preferences~\citep{Ingraham2022Role}. 
Similarly, the algorithm may be useful in human-robot interactions where the robot seeks to assist the human~\citep{li2016framework,nikolaidis2017game}, as previously these applications required solving a costly inverse problem~\citep{ng2000algorithms, li2019differential}.
We evaluate the performance of this algorithm in extensive human subjects experiments, demonstrating consistent convergence to the human's cost minimum.

\begin{figure}[t]
    \begin{minipage}{0.5\linewidth}
    \begin{algorithm}[H]
        \caption{$1 \times 1$ Experiments}\label{alg:exp1x1}
        \SetKwInOut{Require}{Require}
        \Require {Initialize $L$, $\Delta$, $\alpha$, $\wh{h}^*[0]$, $\wh{m}^*[0]$}
        
        \For{$k=0, 1, \dots, K-1$}{
            $\left(h', m'\right) \gets \texttt{trial}\left(L, \wh{h}^*[k], \wh{m}^*[k]\right)$\;
    
            $\left(h'', m''\right)\gets\texttt{trial}\left(L + \Delta, \wh{h}^*[k], \wh{m}^*[k]\right)$\;
            
            $\wh{h}^*[k+1] \gets h'$\;
            
            $\wh{m}^*[k+1] \gets \wh{m}^*[k] + \alpha \left(m'' - \wh{m}^*[k]\right)$\; 
        }
        \SetKwFunction{FMain}{trial}
        \SetKwProg{Fn}{Function}{:}{}
        \Fn{\FMain{$\widetilde{L}, \widetilde{h}^*, \widetilde{m}^*$}}{
            
            \For{$t=0, 1, \dots, T-1$}{
                $h[t] \gets \texttt{get\_manual\_input}\left(t\right)$\;
                
                $m[t] \gets \widetilde{L}\left(h[t] - \widetilde{h}^*\right) + \widetilde{m}^*$\;
                
                $\texttt{display\_cost}\left(c\left(h[t], m[t]\right)\right)$\;
            }
        }
        \KwRet mean of last $t$ iterations of $h$ and $m$\;
    \end{algorithm}
    \end{minipage}
    \begin{minipage}{0.5\linewidth}
        \vspace{12pt}
        \includegraphics[width=1\linewidth]{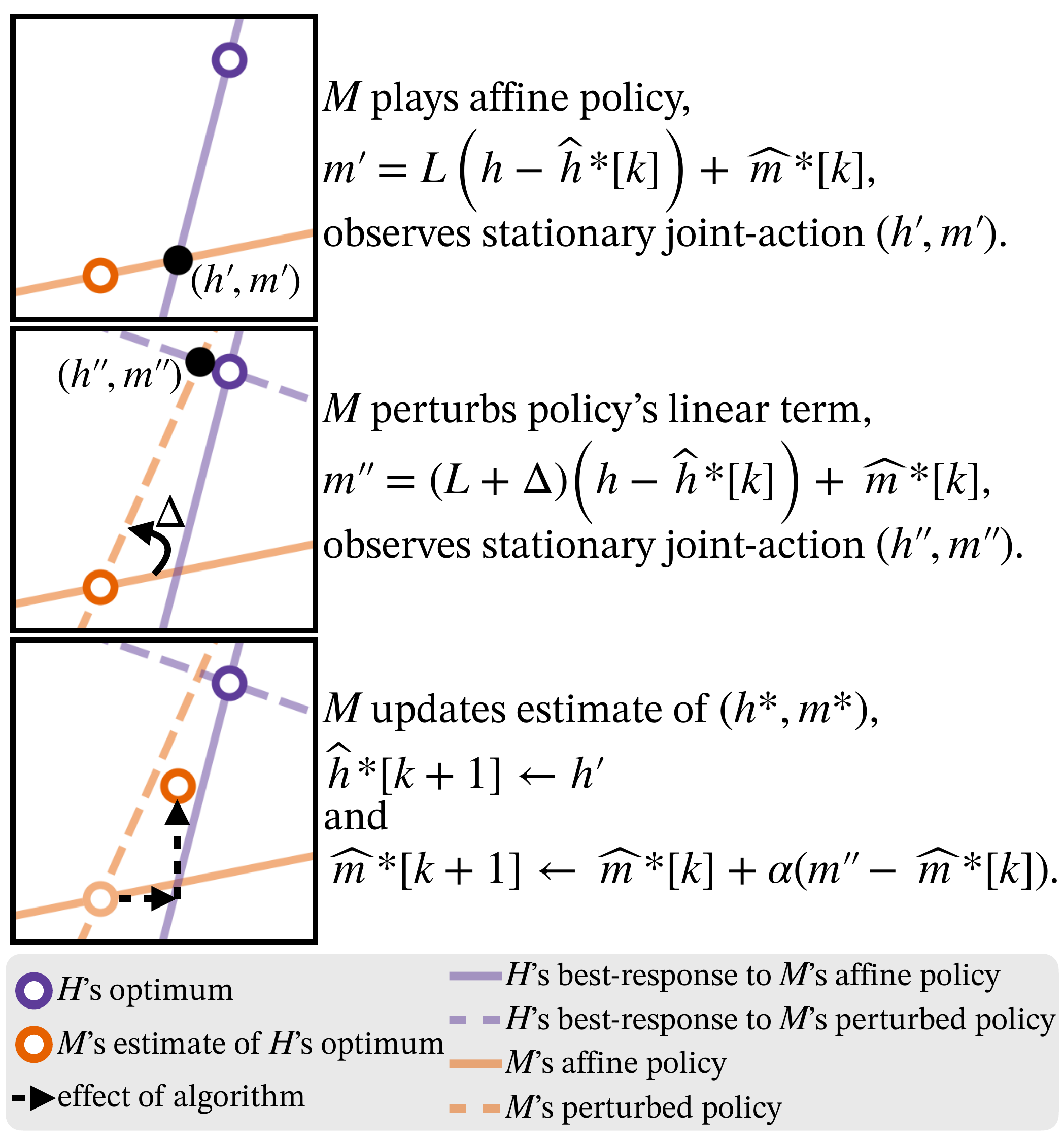}
    \end{minipage}

    \centering
    \label{fig:enter-label}
\end{figure}

\section{Machine's Learning Algorithm}\label{Machine learning algorithm}

We consider a repeated game played between two agents: human $H$ and learning algorithm $M$. 
Both agents seek to minimize the cost function $c:\mc H \times \mc M \to \mathbb{R}$ by individually selecting actions $h\in\mc H\subseteq \R^{d_H}$, $m\in\mc M\subseteq \R^{d_M}$.
However, there is an asymmetry in the information set: we let $H$ have perfect knowledge of $c$, whereas $M$ has no knowledge of $c$.
Finally, we assume that $H$ chooses its action to minimize its cost subject to the constraint imposed by $M$'s current action or policy.
It is $M$'s goal to infer the location of the minimum $\left(h^*,m^*\right)$ of $c$ based on $H$'s responses.
In our experiments, 
$M$ updated its estimate $\left(\wh{h}^*,\wh{m}^*\right)$ of $\left(h^*,m^*\right)$ by observing the human's responses to the affine policy $m = L (h - \wh{h}^*) + \wh{m}^*$. 
We studied the simple quadratic cost
\begin{equation}
    \begin{aligned}
        c(h,m) =
        & \frac{1}{2}h^\top h 
        + \frac{1}{2}m^\top m
    \end{aligned}
    \label{eq:human-cost}
\end{equation}
and let $\left(h^*,m^*\right) = (0,0)$ denote the minimum of $c$.
Algorithm~\ref{alg:exp1x1} specifies the computational procedure and illustrates its three key steps. 

We have previously found in repeated games that people can rapidly play their best-response to machine actions \citep{chasnov2023human,isa2024effect}.
This best-response takes the form
\begin{equation}
    \begin{aligned}
        %BR_H &= 
        \underset{h}{\arg\min}\, c\left(h,L\left(h-\wh{h}^*\right)+\wh{m}^*\right)
        = \left(I + L^\top L\right)^{-1} \left(L^\top L \wh{h}^{* \top} - L^\top \wh{m}^{* \top}\right)
    \end{aligned}
    \label{BR_H}
\end{equation}
when the machine plays the affine policy $m = L \left(h - \wh{h}^*\right) + \wh{m}^*$.
Combining this expression with the formulas from the learning algorithm defines a discrete-time linear system,
\begin{equation}
    \begin{aligned}
        &\left[\begin{matrix}
            \wh{h}^*_{+} \\
            \wh{m}^*_{+}
        \end{matrix}\right]
        &= \left[\begin{matrix}
            (I + L^\top L)^{-1}L^\top L &  - (I + L^\top L)^{-1} L^\top\\
            \alpha L_\Delta (I + L_\Delta^\top L_\Delta)^{-1}L_\Delta^\top L_\Delta - \alpha L_\Delta & I - \alpha L_\Delta(I + L_\Delta^\top L_\Delta)^{-1}L_\Delta^\top
        \end{matrix}\right] \left[\begin{matrix}
            \wh{h}^* \\
            \wh{m}^*
        \end{matrix}\right]
    \end{aligned}
    \label{discrete-time linear system}
\end{equation}
where $L$ is the linear term in $M$'s affine policy and $L_\Delta = L+\Delta$ is the perturbed linear term.
We use this linear system to simulate the human-machine interaction and assess convergence to to $c$'s minimum $(h^*,m^*)$.

\section{Methods}

\begin{figure}[t]
    \center
    \begin{minipage}[c]{0.6\linewidth}
      \includegraphics[width=\textwidth]{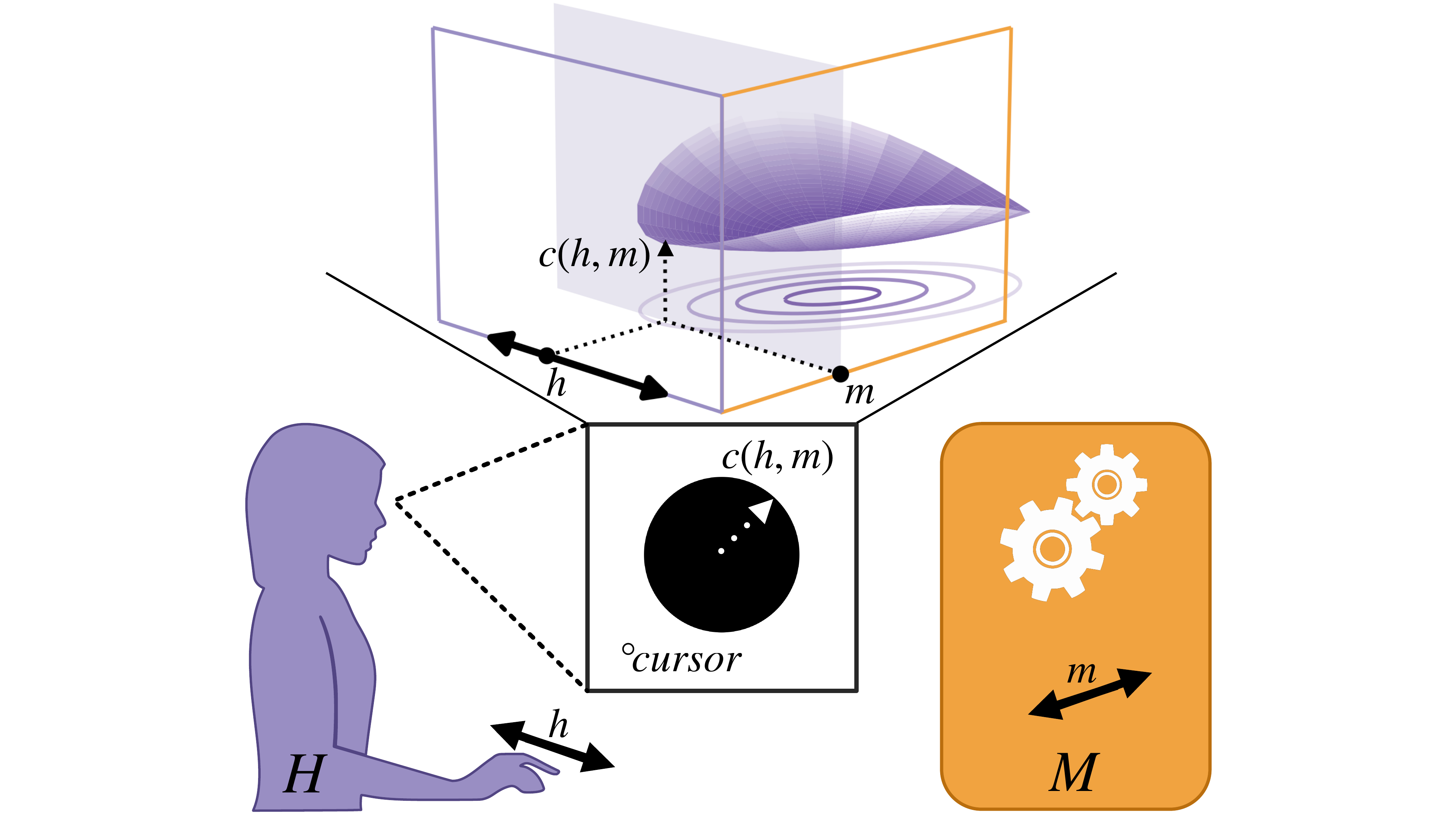}
    \end{minipage}
    \begin{minipage}[c]{0.3\linewidth}
      \caption{Human participant $H$ provides manual input $h$ to keep the black circle on a computer screen as small as possible while a learning algorithm $M$ determines input $m$.  The radius of the circle represents the instantaneous value of a prescribed cost $c(h,m)$.}
      \label{fig:setup}
    \end{minipage}
\end{figure}

We conducted four experiments corresponding to different pairings of action space dimensions for the human and machine agents. We denote these pairings as $d_H \times d_M$ where $d_H$ is the dimension of the human's action and $d_M$ is the dimension of the machine's action. The four cases we tested were $1 \times 1$, $1 \times 2$, $2 \times 1$, and $2 \times 2$. 
Figure~\ref{fig:setup} provides a conceptual overview of the game. 
The instantaneous cost $c(h[t],m[t])$ at time $t$ was continuously rendered to participants during the duration of trials as the radius of a circle on a computer display. 
This method of \emph{prescribing} the cost function to the human was previously tested in~\citep{chasnov2023human,isa2024effect}.
Before and throughout each trial, participants were instructed to ``keep this circle as small as possible''.

\subsection{Participant Population} 
All human subjects were recruited from the crowd-sourcing research platform \emph{Prolific}~\citep{palan2018prolific}. 
Participants had no prior experience with the game and were selected from the standard sample available on the Prolific platform. 

\subsection{Human Input} 
Participants provided manual input via a computer mouse to continuously determine the value of $h$. For the $1 \times 1$ and $1 \times 2$ experiments only the horizontal position of cursor was used to determine the value of a one-dimensional input action $h\in[-1,1]\subset\mathbb{R}$. For the $2 \times 1$ and $2 \times 2$ experiments, participants provided both horizontal and vertical manual input to determine the value of a two-dimensional input action $h\in[-1,1]^2\subset\mathbb{R}^2$.
To ensure that the location of the joint-actions corresponding to the human optimum was not at the center of the screen, the participant’s playable input actions were translated to place the location of $h^*$ to be $\frac{1}{8}$ the width or height of the game screen away from center.

To help prevent human participants from memorizing the location of the game equilibria, we included a ``mirroring'' effect. For the $1 \times 1$ and $1 \times 2$ experiments, a variable $s$ was chosen uniformly at random from $\{-1, +1\}$ at the beginning of each trial and the map $h[t] \to s h[t]$ was applied to the participant's manual input for the duration of the trial. For the $2 \times 1$ and $2 \times 2$ experiments, where the human had two-input actions, the map $h_{i}[t] \to s_i h_{i}[t]$ was applied to the participant’s manual input for the duration of the trial, where $i$ denotes the horizontal ($i=1$) or vertical ($i=2$) input. When the variable's value was $-1$, this had the effect of applying a ``mirror'' symmetry to the input.

\subsection{Experiment Initialization} 
For the $1 \times 1$ experiment, we tested 8 different initializations for the machine's estimate of $c$'s minimum. 
We collected data from 10 naive human subjects for each of the 8 initialization points, totaling 80 subjects for the $1 \times 1$ experiment. 
For the $1 \times 2$, $2 \times 1$, and $2 \times 2$ experiments, the machine's initial estimate of $c$'s minimum was sampled uniformly at random from a ball of radius 0.65 centered at $\left(h^*,m^*\right)$, ensuring that the machine's initial estimate is sufficiently far away from the optimum $\left(h^*,m^*\right)$. 
We collected data from 20 participants for each of the $1 \times 2$, $2 \times 1$, and $2 \times 2$ versions of the game. 
Each subject only participated in one of the experiments.

\subsection{Protocol} 
The $1 \times 1$ experiment follow the protocol in Algorithm~\ref{alg:exp1x1}. Participants repeated the game for 10 iterations, $K$, containing 2 trials per iteration, $\operatorname{trial}\left(L+\Delta,\wh{h}^*[k],\wh{m}^*[k]\right)$ and $\operatorname{trial}\left(L,\wh{h}^*[k],\wh{m}^*[k]\right)$. 
Each trial was 10 seconds long and returned the mean action $(h,m)$ of the last 5 seconds of the trial. At the beginning of each $1 \times 1$ experiment, participants were given one of eight predetermined initialization points for the machine's initial estimate of the human optimum, $(h^*,m^*)$ spaced equally around a circle of radius $0.65$ as shown in Figure~\ref{fig:exp4_1x1}a.

For the $1 \times 2$ and $2 \times 1$ experiment participants repeated the game for 10 iterations ($K$) containing 3 trials per iteration (2 trials for perturbing each element of the $L$ matrix and 1 trial for a non-perturbed $L$ matrix). For the $2 \times 2$ experiment participants repeated the game for 10 iterations ($K$) containing 5 trials per iteration (4 trials for perturbing each element of the $L$ matrix and 1 trial for a non-perturbed $L$ matrix). 
For the  $1 \times 2$ experiment, each trial was 10 seconds long and returned the mean action $(h,m)$ of the last 5 seconds of the trial. For the $2 \times 1$ and $2 \times 2$ experiment, each trial was 25 seconds long and returned the mean action $(h,m)$ of the last 5 seconds of the trial; the increased duration was chosen because subjects found the 2-dimensional minimization problem more challenging. 

Each experiment contains attention check trials in the beginning (before iteration 1), middle (after iteration 5), and end (after iteration 10) of the experiment. Each attention check trial lasted the same amount of time, had the same controls, and same instructions as a normal trial in the experiment. To allow for different optimum action locations on the player's game screen, the optimum action for these attention check trials are placed randomly $\pm \frac{1}{8}$ the width of the game screen away from the center in the horizontal direction for the $1 \times 1$ and $1 \times 2$ experiments and $\pm \frac{1}{8}$ the width/height of the game screen away from the center in the horizontal and vertical direction for the $2 \times 1$ and $2 \times 2$ experiments. 
When participants complete an attention check trial with a mean of the last 5 seconds, selected action within $\pm \frac{1}{8}$ the width/height of the game screen away from the optimum action, the participant is moved forward in the experiment. If participants fail to meet this condition, an attention check trial is repeated. Participants are given 5 attempts to pass the attention check trials before being screened out of the experiment.

\subsection{Data Collection} 
Data for all experiments were collected at a rate of 60 samples per second. 
The data consisted of the time samples $t$, human input $h[t]$, machine input $m[t]$, cost $c(h[t],m[t])$, machine's current estimate $\wh{h}^*[t]$ of $h^*$ and current estimate $\wh{m}^*[t]$ of $m^*$, and cost $c\left(\wh{h}^*[t],\wh{m}^*[t]\right)$.
Different numbers of samples were collected for the $1 \times 1$, $1 \times 2$, $2 \times 1$, and $2 \times 2$ experiments corresponding to their differing trial durations.

\subsection{Simulations}
We implemented simulations for all versions of our human subjects experiments based on the linear system derived in Section~2. 
All simulations ran for 10 iterations and the cost function and algorithm parameters in the simulations are the same as those used in the human subject experiments: the cost function was given by Equation~\ref{eq:human-cost} and the algorithm parameters $\Delta = 1$, $\alpha = 1$, and $L = 0$.

\begin{figure*}[t]
    \centering {\includegraphics[width=1.00\linewidth]{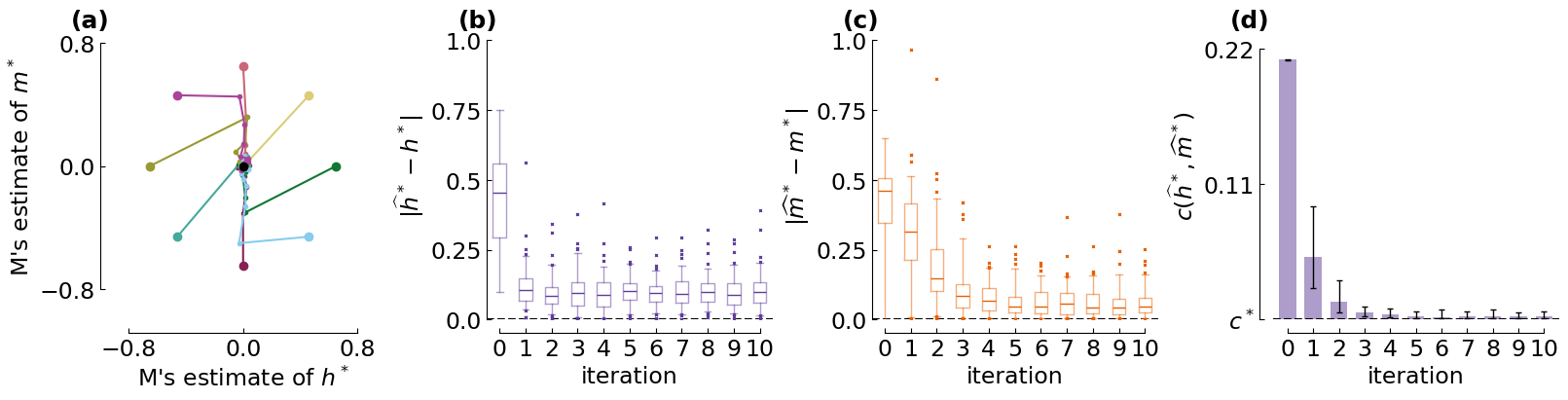}}
    \caption{$1 \times 1$ Experiment ($n = 80$, $n = 10$ per initialization point): (a) $M$’s median estimate of $h^*$ and $m^*$ over iterations for each initialization point. (b) Distributions of L-1 error of $M$’s estimate of $h^*$; box-and-whiskers plot showing 5th, 25th, 50th, 75th, and 95th percentiles. (c) Distributions L-1 error of $M$’s estimate of $m^*$; box-and-whiskers plot showing same as (b). (d) Cost distributions; bar plots with quartiles. The total L-1 error distributions of $M$’s estimate of the optimum $(h^*,m^*)$ can be obtained by adding the L-1 errors from (b) and (c).
    }
    \label{fig:exp4_1x1}
\end{figure*}

\begin{figure*}[t]
    \center
    {\includegraphics[width=1.00\linewidth]{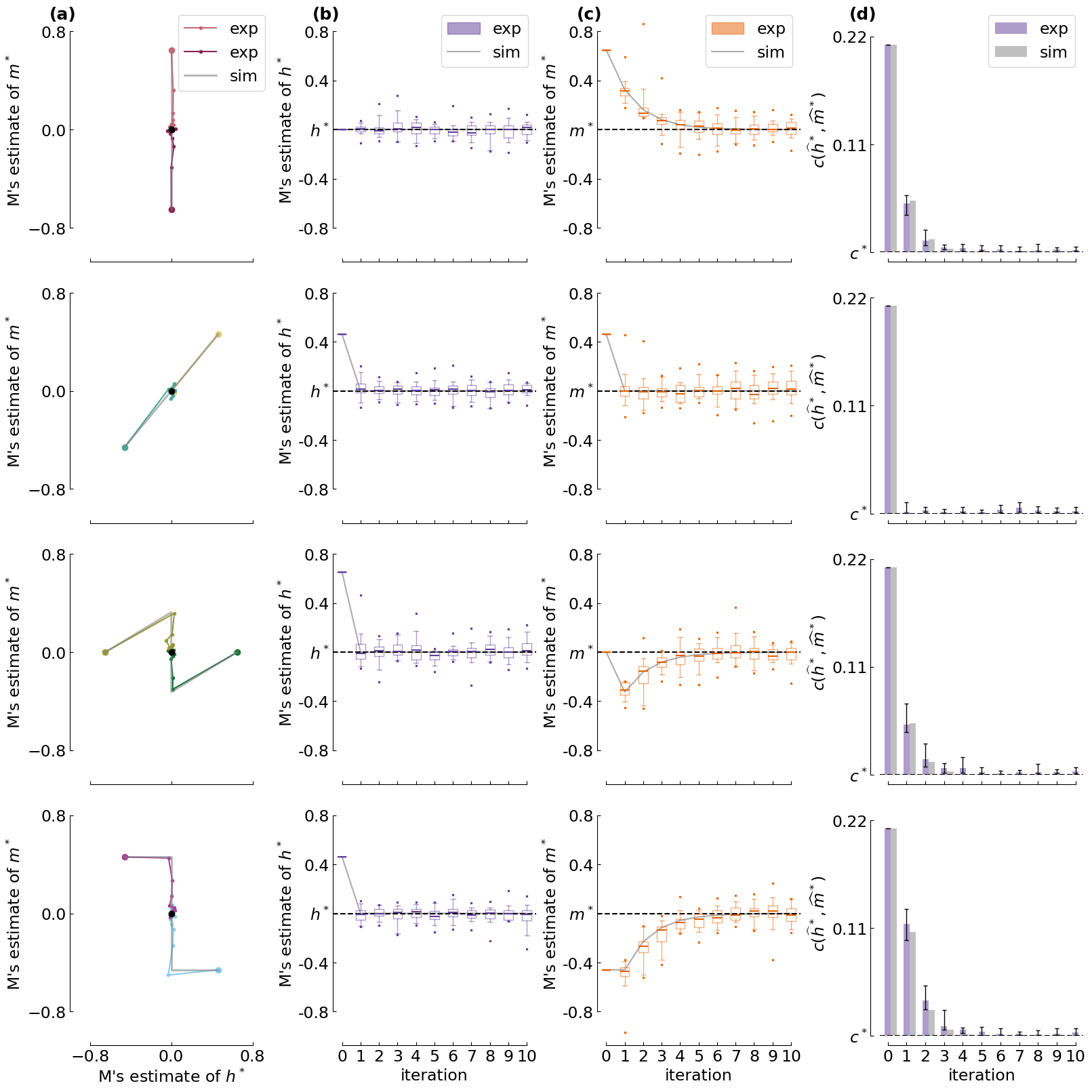}}
    \caption{$1 \times 1$ Simulation vs Experiment ($n = 80$; $n=10$ per initialization point): Grey lines correspond to simulation data. (a) $M$’s median estimate of $h^*$ and $m^*$ over iterations for each initialization point. (b) Distributions of $M$’s estimate of $h^*$; box-and-whiskers plot showing 5th, 25th, 50th, 75th, and 95th percentiles. (c) Distributions $M$’s estimate of $m^*$; box-and-whiskers plot showing same as (b). (d) Cost distributions; bar plots with quartiles.}
    \label{fig:simvexp}
\end{figure*}

\begin{figure*}[t]
    \centering {\includegraphics[width=.88\linewidth]{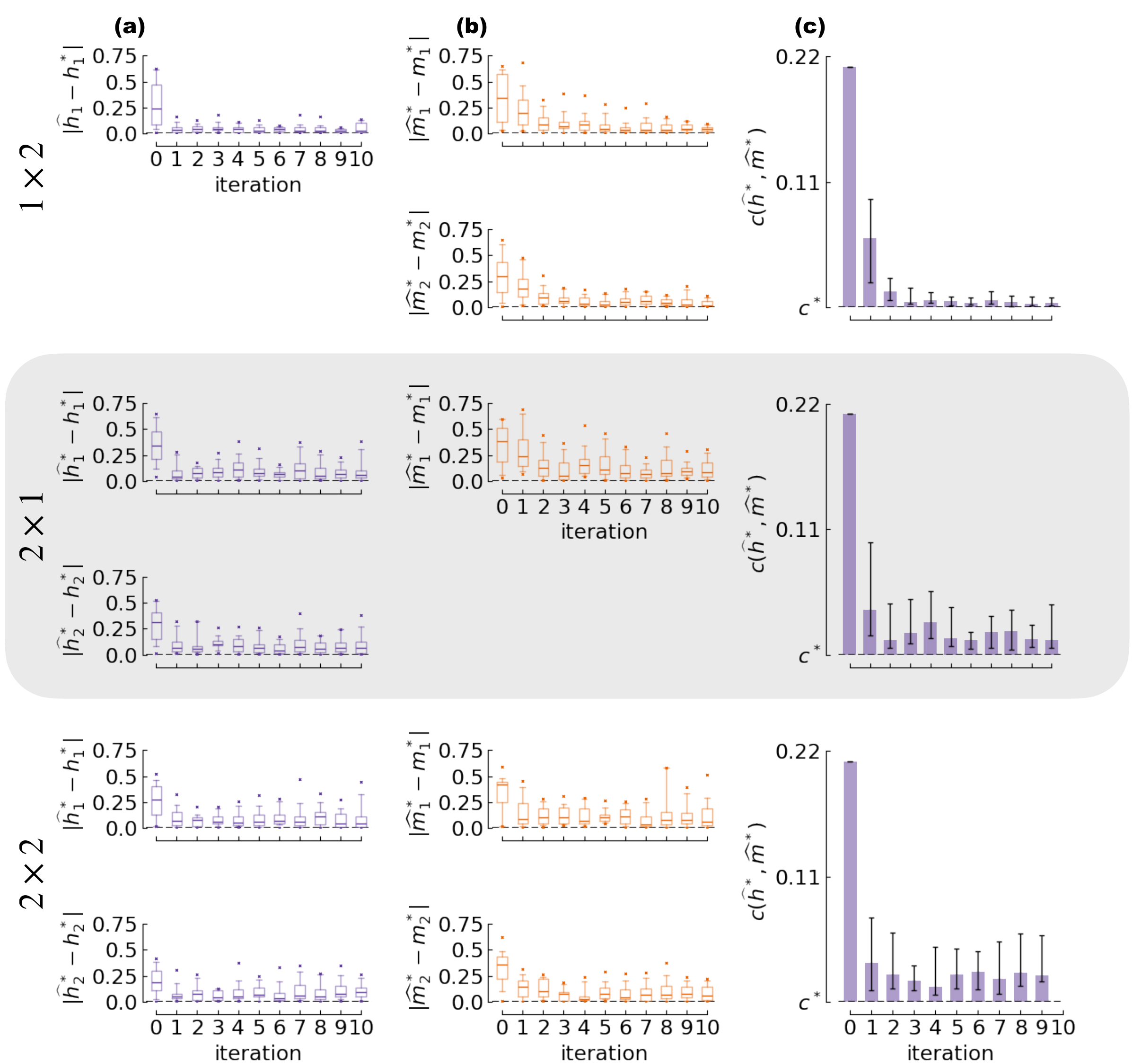}}
    \caption{$1 \times 2$, $2 \times 1$, $2 \times 2$ Experiments ($n = 20$ each): (a) Distributions of L-1 error of $M$’s estimate of $h^*$; box-and-whiskers plot showing 5th, 25th, 50th, 75th, and 95th percentiles. (b) Distributions of L-1 error of $M$’s estimate of $m^*$; box-and-whiskers plot showing same as (a). (c) Cost distributions; bar plots with quartiles.}
    \label{fig:exp_1x2_2x1_2x2}
\end{figure*}

\section{Results}

\subsection{Convergence to $(h^*,m^*)$ in the $1\times1$ game}
Figure~\ref{fig:exp4_1x1} shows summary results for the $1 \times 1$ game. 
From each of 8 symmetrically-arranged initialization points in (a), the machine's estimate $\left(\wh{h}^*,\wh{m}^*\right)$ rapidly converges to the human's cost minimum $\left(h^*,m^*\right)$ in (b,c), leading to corresponding convergence to the minimal value of the cost $c^* = 0$ in (d).
These results are split apart in Figure~\ref{fig:simvexp} into symmetric pairs of initialization points, whose data are combined by applying mirror symmetry across the line bisecting the two points in (a).
The advantage of visualizing the data in this way is that we can observe the signed convergence of the machine estimates to their optimal values in (b,c).
Simulation data is overlaid on the experiment data, showing excellent agreement between theory and experiment.

\subsection{Convergence to $(h^*,m^*)$ in the $1\times2$, $2\times1$, and $2\times2$ games}
Figure~\ref{fig:exp_1x2_2x1_2x2} shows data from the $1\times2$, $2\times1$, and $2\times2$ experiments in an analogous layout as Figure~\ref{fig:exp4_1x1}(b,c,d).
We observe consistent convergence of $\left(\wh{h}^*,\wh{m}^*\right)$ to $\left(h^*,m^*\right)$ in all three experiments, and corresponding convergence to the minimizing cost $c^*$.

\section{Discussion}

We showed consistent convergence of our algorithm to the minimum of a cost function prescribed to human subjects across scalar and multidimensional instantiations of the game.
In particular, the distance in action space between the machine's estimate of the human's cost minimum decreased rapidly, yielding a correspondingly quick decrease in the value of the cost toward its minimum.
These experimental results correspond closely with simulations, providing further evidence that our algorithm works as expected.
This machine's learning algorithm enables the human-machine system to converge to the human's cost minimum \emph{without knowledge or measurements of the cost function}.
Results from pilot experiments (not reported here) demonstrated convergence for the quadratic cost functions from~\citep{chasnov2023human,isa2024effect}, suggesting generalizability of our results.

A key limitation of the present work is that the cost function is prescribed to the human subjects by the experimenter.
We adopted this paradigm since it provided the cleanest proof-of-concept, as there is no uncertainty about the cost function the human subjects seek to minimize.
Going forward, we plan to apply our algorithm in more realistic scenarios, a few of which we outline here.
The context where it is best established that humans minimize a measurable cost function is in assisted mobility, where people are known to minimize metabolic cost-of-transport, which can be measured using respirometry~\citep{Felt2015Body,Zhang2017Human,slade2022personalizing,Abram2022General}.
In this context, our algorithm could be compared against state-of-the-art \emph{human-in-the-loop} optimization algorithms that rely on cost measurements.
Another possible application domain is human-robot interaction~\citep{li2016framework,nikolaidis2017game}, though a key challenge there is that the human's cost function is unknown, so the algorithm can be applied but there will not be a ``baseline'' to compare against.
Finally, the algorithm could be tested in the context of personalized AI~\citep{Zhang2020,10.1145/3512930}, e.g. AI-assisted decision makers who optimize human-user trust, which is unknown and varies between individuals. 

\section{Conclusion}
We presented a new learning algorithm that enables a human-machine interaction to converge to the minimum of a shared cost function.
The novelty in our algorithm is that the cost function is only known to the human, leaving the machine to iteratively update its estimate of the minimum solely by observing the human's response to its actions and policies.
In the case where the cost function is quadratic, our algorithm defines an affine discrete-time system, facilitating analysis of convergence.
Future work could further explore this system's dynamics, potentially providing conditions that guarantee convergence or bounding the convergence rate.

\appendix
\section{Additional Algorithms}

In this section, we provide the modified algorithm for the $1 \times 2$, $2 \times 1$, and $2 \times 2$ experiments.

\begin{center}
\begin{minipage}{1\linewidth}
\begin{minipage}{0.49\linewidth}
\begin{algorithm}[H]
    \footnotesize \caption{$1 \times 2$ Experiments}\label{alg:exp1x2}
    \SetKwInOut{Require}{Require}
    \Require {Initialize $L$, $\Delta$, $\alpha$, $\wh{h}^*[0]$, $\wh{m}^*[0]$}
    
    \For{$k=0, 1, \dots, K-1$}{
        $\left(h', m'\right) \gets \texttt{trial}\left(L, \wh{h}^*[k], \wh{m}^*[k]\right)$\;
        
        $\left(h'', m''\right)\gets\texttt{trial}\left(L + \left[\begin{matrix}
            \Delta\\
            0
        \end{matrix}\right], \wh{h}^*[k], \wh{m}^*[k]\right)$\;

        $\left(h''', m'''\right)\gets\texttt{trial}\left(L + \left[\begin{matrix}
            0\\
            \Delta
        \end{matrix}\right], \wh{h}^*[k], \wh{m}^*[k]\right)$\;

        $\wh{h}^*[k+1] \gets h'$\;
        
        $\wh{m}^*[k+1] \gets \wh{m}^*[k] + \alpha \left(m'' + m''' - 2\wh{m}^*[k]\right)$\;
    }
    \SetKwFunction{FMain}{trial}
    \SetKwProg{Fn}{Function}{:}{}
    \Fn{\FMain{$\widetilde{L}, \widetilde{h}^*, \widetilde{m}^*$}}{
        
        \For{$t=0, 1, \dots, T-1$}{
            $h[t] \gets \texttt{get\_manual\_input}(t)$\;
            
            $m[t] \gets \widetilde{L}\left(h[t] - \widetilde{h}^*\right) + \widetilde{m}^*$\;
            
            $\texttt{display\_cost}\left(c\left(h[t], m[t]\right)\right)$\;
        }
    }
    \KwRet mean of last $t$ iterations of $h$ and $m$\;
\end{algorithm}
\end{minipage} \hspace{0.5 em}
\begin{minipage}{0.49\linewidth}
\begin{algorithm}[H]
    \footnotesize \caption{$2 \times 1$ Experiments}\label{alg:exp2x1}
    \SetKwInOut{Require}{Require}
    \Require {Initialize $L$, $\Delta$, $\alpha$, $\wh{h}^*[0]$, $\wh{m}^*[0]$}
    
    \For{$k=0, 1, \dots, K-1$}{
        $\left(h', m'\right) \gets \texttt{trial}\left(L, \wh{h}^*[k], \wh{m}^*[k]\right)$\;
    
        $\left(h'', m''\right)\gets\texttt{trial}\left(L + \left[\begin{matrix}
            \Delta & 0
        \end{matrix}\right], \wh{h}^*[k], \wh{m}^*[k]\right)$\;

        $\left(h''', m''\right)\gets\texttt{trial}\left(L + \left[\begin{matrix}
            0 & \Delta
        \end{matrix}\right], \wh{h}^*[k], \wh{m}^*[k]\right)$\;

        $\wh{h}^*[k+1] \gets h'$\;
        
        $\wh{m}^*[k+1] \gets \wh{m}^*[k] + \alpha \left(m'' + m''' - 2\wh{m}^*[k]\right)$\;
    }
    \SetKwFunction{FMain}{trial}
    \SetKwProg{Fn}{Function}{:}{}
    \Fn{\FMain{$\widetilde{L}, \widetilde{h}^*, \widetilde{m}^*$}}{
        
        \For{$t=0, 1, \dots, T-1$}{
            $h[t] \gets \texttt{get\_manual\_input}(t)$\;
            
            $m[t] \gets \widetilde{L}\left(h[t] - \widetilde{h}^*\right) + \widetilde{m}^*$\;
            
            $\texttt{display\_cost}\left(c\left(h[t], m[t]\right)\right)$\;
        }
    }
    \KwRet mean of last $t$ iterations of $h$ and $m$\;\
    \vspace{1.3 em}
\end{algorithm}
\end{minipage}
\end{minipage}

\begin{minipage}{0.6\linewidth}
{\scriptsize
\begin{algorithm}[H]
    \footnotesize \caption{$2 \times 2$ Experiments}\label{alg:exp2x2}
    \SetKwInOut{Require}{Require}
    \Require {Initialize $L$, $\Delta$, $\alpha$, $\wh{h}^*[0]$, $\wh{m}^*[0]$}
    
    \For{$k=0, 1, \dots, K-1$}{
        $\left(h', m'\right) \gets \texttt{trial}\left(L, \wh{h}^*[k], \wh{m}^*[k]\right)$\;
    
        $\left(h'', m''\right)\gets\texttt{trial}\left(L + \begin{bmatrix}
            \Delta & 0\\
            0 & 0
        \end{bmatrix}, \wh{h}^*[k], \wh{m}^*[k]\right)$\;

        $\left(h''', m'''\right)\gets\texttt{trial}\left(L + \begin{bmatrix}
            0 & \Delta\\
            0 & 0
        \end{bmatrix}, \wh{h}^*[k], \wh{m}^*[k]\right)$\;

        $\left(h'''', m''''\right)\gets\texttt{trial}\left(L + \begin{bmatrix}
            0 & 0\\
            \Delta & 0
        \end{bmatrix}, \wh{h}^*[k], \wh{m}^*[k]\right)$\;

        $\left(h''''', m'''''\right)\gets\texttt{trial}\left(L + \begin{bmatrix}
            0 & 0\\
            0 & \Delta
        \end{bmatrix}, \wh{h}^*[k], \wh{m}^*[k]\right)$\;

        $\wh{h}^*[k+1] \gets h'$\;
        
        $\wh{m}^*[k+1] \gets \wh{m}^*[k] + \alpha \left(m'' + m''' + m''''  + m''''' - 4\wh{m}^*[k]\right)$\;
    }
    \SetKwFunction{FMain}{trial}
    \SetKwProg{Fn}{Function}{:}{}
    \Fn{\FMain{$\widetilde{L}, \widetilde{h}^*, \widetilde{m}^*$}}{
        
        \For{$t=0, 1, \dots, T-1$}{
            $h[t] \gets \texttt{get\_manual\_input}(t)$\;
            
            $m[t] \gets \widetilde{L}\left(h[t] - \widetilde{h}^*\right) + \widetilde{m}^*$\;
            
            $\texttt{display\_cost}\left(c\left(h[t], m[t]\right)\right)$\;
        }
    }
    \KwRet mean of last $t$ iterations of $h$ and $m$\;
\end{algorithm}
}
\end{minipage}
\end{center}

\newpage
\bibliography{main}

\begin{thebibliography}{28}
\providecommand{\natexlab}[1]{#1}
\providecommand{\url}[1]{\texttt{#1}}
\expandafter\ifx\csname urlstyle\endcsname\relax
  \providecommand{\doi}[1]{doi: #1}\else
  \providecommand{\doi}{doi: \begingroup \urlstyle{rm}\Url}\fi

\bibitem[Abram et~al.(2022)Abram, Poggensee, S{\'a}nchez, Simha, Finley, Collins, and Donelan]{Abram2022General}
Sabrina~J. Abram, Katherine~L. Poggensee, Natalia S{\'a}nchez, Surabhi~N. Simha, James~M. Finley, Steven~H. Collins, and J.~Maxwell Donelan.
\newblock General {V}ariability {L}eads to {S}pecific {A}daptation {T}oward {O}ptimal {M}ovement {P}olicies.
\newblock \emph{Current Biology}, 32\penalty0 (10):\penalty0 2222--2232.e5, 2022.
\newblock \doi{10.1016/j.cub.2022.04.015}.

\bibitem[Cannan and Hu(2011)]{cannan2011human}
James Cannan and Huosheng Hu.
\newblock Human-{M}achine {I}nteraction ({HMI}): A {S}urvey.
\newblock \emph{University of Essex}, 27:\penalty0 46--64, 2011.

\bibitem[Cao et~al.(2020)Cao, Hu, Kiang, and Hong]{cao2020portfolio}
Mukun Cao, Qing Hu, Melody~Y. Kiang, and Hong Hong.
\newblock A {P}ortfolio {S}trategy {D}esign for {H}uman-{C}omputer {N}egotiations in e-{R}etail.
\newblock \emph{International Journal of Electronic Commerce}, 24\penalty0 (3):\penalty0 305--337, 2020.
\newblock \doi{10.1080/10864415.2020.1767428}.

\bibitem[Chasnov et~al.(2023)Chasnov, Ratliff, and Burden]{chasnov2023human}
Benjamin~J. Chasnov, Lillian~J. Ratliff, and Samuel~A. Burden.
\newblock Human adaptation to adaptive machines converges to game-theoretic equilibria, 2023.
\newblock URL \url{https://arxiv.org/abs/2305.01124}.

\bibitem[Crandall et~al.(2018)Crandall, Oudah, Tennom, Ishowo-Oloko, Abdallah, Bonnefon, Cebrian, Shariff, Goodrich, and Rahwan]{crandall2018cooperating}
Jacob~W Crandall, Mayada Oudah, Tennom, Fatimah Ishowo-Oloko, Sherief Abdallah, Jean-Fran{\c{c}}ois Bonnefon, Manuel Cebrian, Azim Shariff, Michael~A Goodrich, and Iyad Rahwan.
\newblock Cooperating with {M}achines.
\newblock \emph{Nature Communications}, 9, 2018.
\newblock \doi{10.1038/s41467-017-02597-8}.

\bibitem[De~Santis(2021)]{DeSantis2021}
Dalia De~Santis.
\newblock A {F}ramework for {O}ptimizing {C}o-adaptation in {B}ody-{M}achine {I}nterfaces.
\newblock \emph{Frontiers in Neurorobotics}, 15, 2021.
\newblock \doi{10.3389/fnbot.2021.662181}.

\bibitem[Felt et~al.(2015)Felt, Selinger, Donelan, and Remy]{Felt2015Body}
Wyatt Felt, Jessica~C. Selinger, J.~Maxwell Donelan, and C.~David Remy.
\newblock ``{B}ody-{I}n-{T}he-{L}oop'': {O}ptimizing {D}evice {P}arameters {U}sing {M}easures of {I}nstantaneous {E}nergetic {C}ost.
\newblock \emph{PLOS ONE}, 10\penalty0 (8):\penalty0 1--21, 08 2015.
\newblock \doi{10.1371/journal.pone.0135342}.

\bibitem[Gorecky et~al.(2014)Gorecky, Schmitt, Loskyll, and Zühlke]{Gorecky2014Human}
Dominic Gorecky, Mathias Schmitt, Matthias Loskyll, and Detlef Zühlke.
\newblock Human-machine-interaction in the industry 4.0 era.
\newblock In \emph{IEEE International Conference on Industrial Informatics (INDIN)}, pages 289--294, 2014.
\newblock \doi{10.1109/INDIN.2014.6945523}.

\bibitem[Hoc(2000)]{Hoc2000From}
Jean-Michel Hoc.
\newblock From human-machine interaction to human-machine cooperation.
\newblock \emph{Ergonomics}, 43\penalty0 (7):\penalty0 833--843, 2000.
\newblock \doi{10.1080/001401300409044}.

\bibitem[Ingraham et~al.(2022)Ingraham, Remy, and Rouse]{Ingraham2022Role}
Kimberly~A. Ingraham, C.~David Remy, and Elliott~J. Rouse.
\newblock The role of user preference in the customized control of robotic exoskeletons.
\newblock \emph{Science Robotics}, 7\penalty0 (64):\penalty0 eabj3487, 2022.
\newblock \doi{10.1126/scirobotics.abj3487}.

\bibitem[Isa et~al.(2024)Isa, Wu, Wang, Zhang, Burden, Ratliff, and Chasnov]{isa2024effect}
Jason~T. Isa, Bohan Wu, Qirui Wang, Yilin Zhang, Samuel~A. Burden, Lillian~J. Ratliff, and Benjamin~J. Chasnov.
\newblock Effect of {A}daptation {R}ate and {C}ost {D}isplay in a {H}uman-{AI} {I}nteraction {G}ame, 2024.
\newblock URL \url{https://arxiv.org/abs/2408.14640}.

\bibitem[Kun(2018)]{Kun2018Human}
Andrew~L. Kun.
\newblock Human-{M}achine {I}nteraction for {V}ehicles: {R}eview and {O}utlook.
\newblock \emph{Foundations and Trends{\textregistered} in Human--Computer Interaction}, 11\penalty0 (4):\penalty0 201--293, 2018.
\newblock \doi{10.1561/1100000069}.

\bibitem[Li et~al.(2016)Li, Tee, Yan, Chan, and Wu]{li2016framework}
Yanan Li, Keng~Peng Tee, Rui Yan, Wei~Liang Chan, and Yan Wu.
\newblock A {F}ramework of {H}uman–{R}obot {C}oordination {B}ased on {G}ame {T}heory and {P}olicy {I}teration.
\newblock \emph{IEEE Transactions on Robotics}, 32\penalty0 (6):\penalty0 1408--1418, 2016.
\newblock \doi{10.1109/TRO.2016.2597322}.

\bibitem[Li et~al.(2019)Li, Carboni, Gonzalez, Campolo, and Burdet]{li2019differential}
Yanan Li, Gerolamo Carboni, Franck Gonzalez, Domenico Campolo, and Etienne Burdet.
\newblock Differential game theory for versatile physical human-robot interaction.
\newblock \emph{Nature Machine Intelligence}, 1\penalty0 (1):\penalty0 36--43, 2019.
\newblock \doi{10.1038/s42256-018-0010-3}.

\bibitem[March(2021)]{march2021strategic}
Christoph March.
\newblock Strategic interactions between humans and artificial intelligence: {L}essons from experiments with computer players.
\newblock \emph{Journal of Economic Psychology}, 87:\penalty0 102426, 2021.
\newblock \doi{10.1016/j.joep.2021.102426}.

\bibitem[Merel et~al.(2013)Merel, Fox, Jebara, and Paninski]{Merel2013-jd}
Josh~S Merel, Roy Fox, Tony Jebara, and Liam Paninski.
\newblock A multi-agent control framework for co-adaptation in brain-computer interfaces.
\newblock In \emph{Advances in Neural Information Processing Systems}, volume~26. Curran Associates, Inc., 2013.
\newblock URL \url{https://proceedings.neurips.cc/paper_files/paper/2013/file/286674e3082feb7e5afb92777e48821f-Paper.pdf}.

\bibitem[Neumann and Morgenstern(1944)]{VonNeumann1944}
John~Von Neumann and Oskar Morgenstern.
\newblock \emph{Theory of Games and Economic Behavior}.
\newblock Princeton University Press, 1944.

\bibitem[Ng and Russell(2000)]{ng2000algorithms}
Andrew~Y. Ng and Stuart~J. Russell.
\newblock Algorithms for {I}nverse {R}einforcement {L}earning.
\newblock In \emph{Proceedings of the Seventeenth International Conference on Machine Learning}, pages 663--670. Morgan Kaufmann Publishers Inc., 2000.

\bibitem[Nikolaidis et~al.(2017)Nikolaidis, Nath, Procaccia, and Srinivasa]{nikolaidis2017game}
Stefanos Nikolaidis, Swaprava Nath, Ariel~D. Procaccia, and Siddhartha Srinivasa.
\newblock Game-{T}heoretic {M}odeling of {H}uman {A}daptation in {H}uman-{R}obot {C}ollaboration.
\newblock In \emph{ACM/IEEE International Conference on Human-Robot Interaction}, pages 323--331. Association for Computing Machinery, 2017.
\newblock \doi{10.1145/2909824.3020253}.

\bibitem[Palan and Schitter(2018)]{palan2018prolific}
Stefan Palan and Christian Schitter.
\newblock Prolific.ac---{A} subject pool for online experiments.
\newblock \emph{Journal of Behavioral and Experimental Finance}, 17:\penalty0 22--27, 2018.
\newblock \doi{10.1016/j.jbef.2017.12.004}.

\bibitem[Perdikis and Millan(2020)]{perdikis2020brain}
Serafeim Perdikis and Jose del~R. Millan.
\newblock Brain-machine interfaces: A tale of two learners.
\newblock \emph{IEEE Systems, Man, and Cybernetics Magazine}, 6\penalty0 (3):\penalty0 12--19, 2020.
\newblock \doi{10.1109/MSMC.2019.2958200}.

\bibitem[Rastogi et~al.(2022)Rastogi, Zhang, Wei, Varshney, Dhurandhar, and Tomsett]{10.1145/3512930}
Charvi Rastogi, Yunfeng Zhang, Dennis Wei, Kush~R. Varshney, Amit Dhurandhar, and Richard Tomsett.
\newblock Deciding {F}ast and {S}low: {T}he {R}ole of {C}ognitive {B}iases in {AI}-assisted {D}ecision-making.
\newblock \emph{Proc. ACM Hum.-Comput. Interact.}, 6\penalty0 (CSCW1), 2022.
\newblock \doi{10.1145/3512930}.

\bibitem[Recht(2019)]{Recht2019}
Benjamin Recht.
\newblock A {T}our of {R}einforcement {L}earning: {T}he {V}iew from {C}ontinuous {C}ontrol.
\newblock \emph{Annual Review of Control, Robotics, and Autonomous Systems}, 2:\penalty0 253--279, 2019.
\newblock \doi{10.1146/annurev-control-053018-023825}.

\bibitem[Simon(1955)]{simmon1955}
Herbert~A. Simon.
\newblock A {B}ehavioral {M}odel of {R}ational {C}hoice.
\newblock \emph{The Quarterly Journal of Economics}, 69\penalty0 (1):\penalty0 99--118, 1955.
\newblock \doi{10.2307/1884852}.

\bibitem[Slade et~al.(2022)Slade, Kochenderfer, Delp, and Collins]{slade2022personalizing}
Patrick Slade, Mykel~J. Kochenderfer, Scott~L. Delp, and Steven~H. Collins.
\newblock Personalizing exoskeleton assistance while walking in the real world.
\newblock \emph{Nature}, 610:\penalty0 277--282, 2022.
\newblock \doi{10.1038/s41586-022-05191-1}.

\bibitem[Todorov and Jordan(2002)]{Todorov2002-sb}
E~Todorov and M~I Jordan.
\newblock {Optimal feedback control as a theory of motor coordination}.
\newblock \emph{Nature neuroscience}, 5\penalty0 (11):\penalty0 1226--1235, 2002.
\newblock \doi{10.1038/nn963}.

\bibitem[Zhang et~al.(2017)Zhang, Fiers, Witte, Jackson, Poggensee, Atkeson, and Collins]{Zhang2017Human}
Juanjuan Zhang, Pieter Fiers, Kirby~A. Witte, Rachel~W. Jackson, Katherine~L. Poggensee, Christopher~G. Atkeson, and Steven~H. Collins.
\newblock Human-in-the-loop optimization of exoskeleton assistance during walking.
\newblock \emph{Science}, 356\penalty0 (6344):\penalty0 1280--1284, 2017.
\newblock \doi{10.1126/science.aal5054}.

\bibitem[Zhang et~al.(2020)Zhang, Liao, and Bellamy]{Zhang2020}
Yunfeng Zhang, Q.~Vera Liao, and Rachel K.~E. Bellamy.
\newblock Effect of confidence and explanation on accuracy and trust calibration in {AI}-assisted decision making.
\newblock In \emph{Proceedings of the 2020 Conference on Fairness, Accountability, and Transparency}, page 295–305. Association for Computing Machinery, 2020.
\newblock \doi{10.1145/3351095.3372852}.

\end{thebibliography}

\end{document}